\documentstyle[12pt]{article}
\makeatletter

\expandafter\ifx\csname mathrm\endcsname\relax\def\mathrm#1{{\rm #1}}\fi

\makeatother

\def\stars{\strut\leaders\hbox{*}\hfill\strut}
\def\starline{\hfil\strut\hfil\hbox to \textwidth {\stars}\hfil}

\makeatletter
\@addtoreset{equation}{section}
\makeatother

\def\beq{\begin{equation}}
\def\eeq{\end{equation}}
\def\beqar{\begin{eqnarray}}
\def\eeqar{\end{eqnarray}}
\def\barr#1{\begin{array}{#1}}
\def\earr{\end{array}}
\def\bfi{\begin{figure}}
\def\efi{\end{figure}}
\def\btab{\begin{table}}
\def\etab{\end{table}}
\def\bce{\begin{center}}
\def\ece{\end{center}}
\def\nn{\nonumber}

\def\text{\textstyle}

\begin{document}

\renewcommand{\thefootnote}{\fnsymbol{footnote}}
\setcounter{footnote}{1}

\hspace*{\fill} BI-TP 95/24\\
\hspace*{\fill} UdeM-GPP-TH-95-29\\
\hspace*{\fill} hep-ph/9506472\\
\hspace*{\fill} June 1995\\
\vspace{1cm}

\begin{center}
\Large{\bf
On Non-standard Couplings among the \\
Electroweak Vector Bosons\footnote{Partially supported by the
EC-network contract CHRX-CT94-0579 and by the BMBF, Bonn, Germany}}
\end{center}

\vspace{1cm}
\begin{center}
{\bf C. Grosse-Knetter\footnote{On leave of absence from the
Universit\"at Bielefeld},}\\
Universit\'e de Montr\'eal, Laboratoire de Physique Nucl\'eaire,\\
C.P. 6128, Montr\'eal, Qu\'ebec, H3C 3J7, Canada\\[5mm]
{\bf I. Kuss\footnote{e-mail: kuss@hrz.uni-bielefeld.de}
and D. Schildknecht}  \\
Fakult\"at f\"ur Physik, Universit\"at Bielefeld, \\
Postfach 10 01 31, D-33501 Bielefeld, Germany \\
\end{center}
\vspace{1cm}

\begin{abstract}
Application of a
Stueckelberg transformation allows one to
connect various Lagrangians which have been independently proposed for
non-standard couplings. We discuss the reduction of the number of independent
parameters in the Lagrangian and compare symmetry arguments with dimensional
arguments.
\end{abstract}

\thispagestyle{empty}
\setcounter{page}{0}

\newpage

\renewcommand{\thefootnote}{\arabic{footnote}}
\setcounter{footnote}{0}

\section{Introduction}

As previously shown \cite{kmss,krs},
reasonable ($SU(2)$ symmetry) constraints, weaker
than the ones embodied in the standard $SU(2)_L \times U(1)_Y$
electroweak theory, allow one to systematically reduce the number of free
parameters compatible with relativistic invariance
of the trilinear and quadrilinear couplings among the
electroweak vector bosons. The results are of theoretical as
well as of practical interest. From the theoretical side they clarify
the connection between non-standard couplings and symmetries.
{}From the practical side, a systematic
reduction by additional symmetry principles of the number of free
parameters allowed by relativistic invariance is mandatory if
reasonable bounds on such parameters are to be obtained by future
measurements, e.g. in $e^+ e^- \rightarrow W^+ W^-$.

In the present note we demonstrate how to systematically extend any
$U(1)_{em}$-gauge-invariant Lagrangian describing
couplings among vector bosons
to become invariant under local $SU(2)_L\times U(1)_Y$ transformations.
The procedure is simple. By applying a Stueckelberg transformation
to the given Lagrangian, local
$SU(2)_L\times U(1)_Y$ invariance becomes manifest via the
introduction of three (unphysical) scalar degrees of freedom
\cite{stueckel,kunimasa,knetterkoe}, which are non-linearly realized.
In a second substitution, one linearizes the theory with respect to the
scalar degrees of freedom by introducing a physical scalar (Higgs boson).

Both the non-linear \cite{appelquist,appelwu} and the linear
\cite{buwy,deru,hagiwara,goure2} realization
of $SU(2)_L\times U(1)_Y$ symmetry
for non-standard couplings among the vector bosons have been given before.
This is not the point of the present paper. The aim of
the present note is twofold. First of all, we show how the different
Lagrangians independently advocated for and discussed in the literature are
{\em connected by a simple transformation} of the vector-boson interactions
via introducing scalar degrees of freedom.
Even though the different Lagrangians are well-known, this simple
interrelation via a Stueckelberg transformation has to the best of our
knowledge never been explicitly presented. Secondly, and as a consequence
of the aforementioned interrelation of the different Lagrangians, it will
become obvious that {\em no additional arguments are gained} concerning
the strategy for reducing the number and nature of the non-standard
interactions of the vector bosons among each other if these interactions are
supplemented by interactions with scalar degrees of freedom.

The auxiliary scalar fields in the non-linear realization allow for a
transition to arbitrary gauges. This is of relevance for loop
calculations. The introduction of a physical Higgs particle
formally improves the degree of divergence of loops calculated within the
theory \cite{hagiwara,hewe}. However, due to the well-known ambiguities
inherently connected with loop calculations in non-renormalizable theories,
the interpretation of the results of such calculations is
quite controversial\footnote{Assuming that non-standard
couplings are generated by one-loop effects of unknown
heavy particles, one should keep in mind that
these one-loop-generated terms must not be
inserted in loops again within a consistent one-loop calculation
\cite{sdcgk}.}.
On the other hand, it may be worthwhile to explore what kind of
Higgs interactions are generated, once one allows for non-standard
couplings.

In section 2, we briefly consider standard
$SU(2)_L \times U(1)_Y$ interactions among vector bosons. This is
necessary for the motivation and a transparent presentation of the
$SU(2)$ symmetry assumptions (local $SU(2)$ as well as the so-called
custodial $SU(2)$, known as $SU(2)_C$) to be employed in section 3 with
respect to the general relativistically invariant Ansatz for
non-standard couplings among the vector bosons. Final
conclusions will be drawn in section 4.

\section{Standard Interactions, non-linearly and linearly realized
\protect\boldmath$SU(2)_L \times U(1)_Y$ symmetry}

We start from the gauge group $SU(2)_L \times U(1)_Y$
\cite{glashow,weinberg}.
Introducing the weak-isospin triplet $W^i_\mu (x)$,
$(i = 1,2,3)$, and a weak hypercharge singlet $B_\mu^0 (x)$, we have
\beq
{\cal L}_{\mathrm{YM}} =
- \frac{1}{2}\,{\mathrm{tr}}\, (W^{\mu\nu} W_{\mu\nu} )
- \frac{1}{2}
\,{\mathrm{tr}}\,(B^{\mu\nu} B_{\mu\nu} )
\label{kin},
\eeq
where
\beqar
W_\mu &=& W_\mu^i \frac{\tau_i}{2}, \qquad
W_{\mu\nu} \,= \,W_{\mu\nu}^i\frac{\tau_i}{2}\;=\;
\partial_\mu W_\nu -
\partial_\nu W_\mu + ig [ W_\mu , W_\nu ],\nn\\
B_\mu &=& B_\mu^0 \frac{\tau_3}{2}, \qquad
B_{\mu\nu} \,=\, B_{\mu\nu}^0\frac{\tau_3}{2}\;=\;
\partial_\mu B_\nu -
\partial_\nu B_\mu .
\label{2.2}
\eeqar
In the absence of a mass term (and even without auxiliary scalar fields),
the Lagrangian (\ref{kin}) is invariant under local $SU(2)_L \times
U(1)_Y$ transformations,
\beqar
W_\mu &\to& SW_\mu S^\dagger - \frac{i}{g} S \partial_\mu S^\dagger
,\qquad S = \exp \left( \frac{i}{2} g\alpha_i \tau_i \right), \qquad
 \alpha_i = \alpha_i (x),\; i=1,2,3,\nn  \\
B_\mu &\to& B_\mu - \partial_\mu \beta\frac{\tau_3}{2}, \qquad
\qquad\qquad\qquad\qquad\qquad\qquad\;\;\;\;\beta = \beta (x),
\label{2.3}
\eeqar
which imply the corresponding transformations of the field strength
tensors,
\beqar
W_{\mu\nu}&\to& S\,W_{\mu\nu}\,S^\dagger,
\nn\\
B_{\mu\nu}&\to& B_{\mu\nu}.
\eeqar
Under local electromagnetic gauge transformations, $U(1)_{em}$,
specified by
\beqar
S&=&\exp\left(\frac{i}{2}\,e\,\chi\tau_3\right),\nn\\
B_\mu&\to& B_\mu-\partial_\mu\chi\frac{\tau_3}{2}\frac{e}{g'},
\label{localem}\eeqar
the fields transform as\footnote{For completeness we note that
$W_{\mu}^\pm\equiv \frac{1}{\sqrt{2}}(W_{\mu}^1
\,\mp\,i\,W_{\mu}^2)$ and
$W_{\mu\nu}^\pm\equiv \frac{1}{\sqrt{2}}(W_{\mu\nu}^1
\,\mp\,i\,W_{\mu\nu}^2)$ with
$W_{\mu\nu}^i=\partial_\mu W_\nu^i -\partial_\nu W_\mu^i -g\,\epsilon_{ijk}
W_\mu^j W_\nu^k$.}
\beq
B_\mu^0 \rightarrow B_\mu^0 - \frac{e}{g^\prime} \partial_\mu \chi,
\qquad\quad B^0_{\mu\nu}\to B^0_{\mu\nu},
\eeq
\beq
W_{\mu}^3 \rightarrow W_{\mu}^3 - \frac{e}{g} \partial_\mu \chi,
\qquad\quad
W_{\mu\nu}^3 \to W_{\mu\nu}^3,
\eeq
and
\beq
W^\pm_\mu \rightarrow \exp \left( \pm i e \chi \right)
W^\pm_\mu,\qquad\quad
W_{\mu\nu}^\pm\to \exp\left( \pm i e \chi \right) W^\pm_{\mu\nu}.
\label{emcharged}
\eeq
A mass term which allows for mixing among neutral bosons and
preserves $U(1)_{em}$ is given by
\beq
{\cal L}_{\mathrm{mass}} = M^2_W \,{\mathrm{tr}}\, \left( W_\mu -
\frac{g^\prime}{g} B_\mu \right)^2.
\label{2.5}
\eeq
Upon diagonalization, (\ref{2.5}) together with (\ref{kin}) yields the
vector-boson part of the Lagrangian of the electroweak standard model
\cite{glashow,weinberg} in its unitary gauge.

The mass term (\ref{2.5}) by assumption has the important property of
being
invariant under global $SU(2)_L$ transformations in the limit of a
decoupling $B_\mu$ field, $g^\prime \rightarrow 0$, i.e., ``intrinsic
$SU(2)$ violation''
\cite{kmss} is excluded for the charged and neutral (unmixed)
masses, all given by $M_{W^\pm} \equiv M_{W^0} \equiv M_W$ in (\ref{2.5}).
This symmetry of the Lagrangian (\ref{kin}), (\ref{2.5}) is also known
as ``custodial'' $SU(2)$ and coincides with the global $SU(2)$ broken by
current mixing employed by Bjorken and Hung and Sakurai \cite{hung}.
It guarantees the validity of Weinberg's mass relation \cite{weinberg}
between the $W^\pm$ and the $Z^0$ masses.

Empirically, by combining the measurements of the $W^\pm$ mass with the
precision data on $e^+ e^- \rightarrow Z_0 \rightarrow$ (leptons, quarks),
one finds a deviation from unity in the ratio $M_{W^0}/M_{W^\pm}$ which is
given by \cite{didi}
\beq
\Delta x^{\mathrm{exp}}
\equiv 1 - \frac{M^2_{W^0}}{M^2_{W^\pm}} = (9.6 \pm 4.7 \pm 0.2)
\cdot 10^{-3}.
\label{2.6}
\eeq
The first error in this result is statistical, while the second one
corresponds to the error in the input value of $\alpha (M^2_Z )^{-1} =
128.87 \pm 0.12$. Even though $\Delta x^{\mathrm{exp}}$ in (\ref{2.6})
deviates from
zero, (\ref{2.6}) provides strong empirical support for $SU(2)_C$
symmetry of the Lagrangian. Indeed, the apparent violation of $SU(2)_C$ in
(\ref{2.6}) can be fully explained as a consequence of the fermion loop
correction dominated by the top-quark loop and given by \cite{didi}
\beq
\Delta x_{\mathrm{ferm}} \simeq 12\cdot 10^{-3}
\;\mbox{for}\;m_t = 180\,\,GeV.
\label{2.7}
\eeq
Accordingly, there is strong motivation
to extend the validity of $SU(2)_C$ to the (non-standard) couplings among
the electroweak vector bosons to be discussed in section 3.

We return to our discussion of the symmetry properties of the
Lagrangian given by (\ref{kin}) and (\ref{2.5}). The mass term as
it stands appears to break local $SU(2)_L \times U(1)_Y$ symmetry. Local
$SU(2)_L \times U(1)_Y$ symmetry becomes manifest, however, upon
introducing auxiliary scalar fields (Goldstone fields) $\varphi_i (x)$
via the (non-Abelian) Stueckelberg transformation
\cite{kunimasa,knetterkoe}
\beq
 W_\mu \rightarrow U^\dagger W_\mu U - \frac{i}{g} U^\dagger
\partial_\mu U,
\qquad B_\mu \rightarrow B_\mu ,
\label{2.8}
\eeq
where
\beq
U \equiv \exp \left( \frac{i}{2} \frac{g}{M_W} \varphi_i \tau_i \right).
\label{2.9}
\eeq
This implies for
the corresponding transformation of the field strength tensors
\beq
 W_{\mu\nu} \rightarrow U^\dagger W_{\mu\nu} U , \qquad
B_{\mu\nu} \to B_{\mu\nu}.
\label{2.14}
\eeq
The substitution (\ref{2.8}) leaves the Yang-Mills term
(\ref{kin}) invariant. Introducing the covariant derivative,
\beq
D_\mu U \equiv \partial_\mu U + ig W_\mu U - i g^\prime U
B_\mu,
\label{2.11}
\eeq
the substitution (\ref{2.8}) results in
\beq
W^\mu-\frac{g'}{g}B^\mu\to -\frac{i}{g}U^\dagger D^\mu U
=\frac{i}{g}(D^\mu U)^\dagger U,
\label{transf}\eeq
and the mass term in (\ref{2.5}) takes the form
\beq
{\cal L}_{\mathrm{mass}} = \frac{M^2_W}{g^2} \, {\mathrm{tr}}\,
\left[ ( D_\mu U)^\dagger
(D^\mu U ) \right].
\label{2.10}
\eeq
The local $SU(2)_L \times U(1)_Y$ transformations (\ref{2.3}) and
\beq
U \rightarrow SU \exp \left( \frac{-i}{2} g^\prime \beta \tau_3 \right)
\label{2.12}
\eeq
assure gauge
invariance of the full electroweak Lagrangian including the
vector-boson mass term. A suitable gauge
fixing condition
yields $U = 1$
\cite{knetterkoe}
and takes us back to the original Lagrangian which is thus
identified as the unitary gauge (U-gauge) of
an $SU(2)_L \times U(1)_Y$ gauge invariant massive vector boson
theory with mixing in the neutral sector.
We note in passing that electromagnetic $U(1)_{em}$ gauge invariance of the
Lagrangian in the original form (\ref{kin}), (\ref{2.5}) is a
prerequisite for local $SU(2)_L \times U(1)_Y$
invariance. An arbitrary $SU(2)_L \times U(1)_Y$ gauge transformation
(\ref{2.3}), (\ref{2.12}) acts as an
electromagnetic gauge transformation on the (U-gauge) Lagrangian (\ref{kin}),
(\ref{2.5}) \cite{knetterkoe}.

In a final step, we linearize in the scalar fields $\varphi_i (x)$ and add
an additional scalar field, the physical Higgs scalar $H(x)$, via the
replacement \cite{knetterkoe,appelquist}
\beqar
U \rightarrow \frac{\sqrt 2}{v} \phi & \equiv &
\frac{1}{\sqrt 2} \frac{g}{M_W}
\phi \nn \\
&=& 1 + \frac{g}{2M_W} \left( H(x) + i\varphi_i (x) \tau_i
\right),
\label{2.13}
\eeqar
and the addition of the Higgs potential terms to the Lagrangian.

We have thus reconstructed \cite{knetterkoe,didi2}
the bosonic sector of the standard electroweak theory in four distinct
steps

\begin{itemize}
\item[i)]
local $SU(2)_L \times U(1)_Y$ symmetry purely in the vector-boson sector
(without scalar fields),
\item[ii)]
custodial $SU(2)$ when introducing vector-boson masses,
\item[iii)]
manifest local $SU(2)_L \times U(1)_Y$ symmetry via three
scalar degrees of freedom within a
non-linear framework, and
\item[iv)]
linearisation and the introduction of a physical scalar (Higgs) particle.
\end{itemize}
The same sequence of steps will now be applied to the most general
Ansatz allowed by relativistic invariance for couplings of vector bosons
among each other.

\section{Non-standard couplings among vector bosons}

Upon diagonalization of the mass term (\ref{2.5}) in the Lagrangian
(\ref{kin}), (\ref{2.5}) via
\beqar
A_\mu&=&{~}c_W B_\mu^0 + s_W W_\mu^3,\nn\\
Z_\mu&=&-s_W B_\mu^0 + c_W W_\mu^3,\label{physfields}
\eeqar
one finds the standard trilinear interactions between
the photon, the neutral $Z$ and the charged $W^\pm$ bosons,
\beqar
{\cal L}_{\mathrm{int, SM}} & = &
-ie [ A_\mu ( W^{-\mu\nu}_{(A)} W^+_\nu - W^{+\mu\nu}_{(A)} W^-_\nu )
+ F_{\mu\nu}^{(A)}
W^{+\mu} W^{-\nu} ] \nn \\
& & -ie \frac{c_W}{s_W} [ Z_\mu (W^{-\mu\nu}_{(A)} W^+_\nu -
W^{+\mu\nu}_{(A)}
 W^-_\nu ) +Z_{\mu\nu}^{(A)} W^{+\mu} W^{-\nu} ],
\label{3.2}
\eeqar
where $W^{\pm\mu\nu}_{(A)}=\partial^\mu W^{\pm\nu}-\partial^\nu
W^{\pm\mu}$,
$F^{\mu\nu}_{(A)}=\partial^\mu A^{\nu}-\partial^\nu A^{\mu}$ and
$Z^{\mu\nu}_{(A)}=\partial^\mu Z^{\nu}-\partial^\nu Z^{\mu}$
are the Abelian field strength tensors.
Here, $e=g\,s_W\,=\,g'\,c_W$ denotes the electromagnetic coupling, where
$s_W^2 \equiv 1 - c_W^2$.
We generalize\footnote{
We restrict ourselves to C and P conserving extensions.} (\ref{3.2})
by allowing for deviations in the relative
magnitude of the two terms making up the photon and the $Z$ interaction,
and by allowing for deviations in the absolute magnitude of the $Z$
interaction.
In addition, Lorentz- as well as C- and P-invariance
allow for a dimension-six term, sometimes called quadrupole term.
Accordingly, we supplement (\ref{3.2}) by the non-standard couplings
\cite{gaemers,billy}
\beqar
{\cal L}_{\mathrm{int, NS}} & = & - i e x_\gamma F_{\mu\nu} W^{+\mu}
W^{-\nu} \nn \\
&  & - i e \delta_Z [ Z_\mu ( W^{-\mu\nu} W^+_\nu - W^{+\mu\nu} W^-_\nu )
+ Z_{\mu\nu} W^{+\mu} W^{-\nu}] \nn \\
&  & - i e x_Z Z_{\mu\nu} W^{+\mu} W^{-\nu} , \nn \\
&  & + i e \frac{y_\gamma}{M^2_{W^\pm}} F_\mu^{~\nu} W_\nu^{-\lambda}
W_\lambda^{+\mu}, \nn \\
& & + i e \frac{y_Z}{M^2_{W^\pm}} Z_\mu^{~\nu} W_\nu^{-\lambda} W_\lambda
^{+\mu},
\label{3.3}
\eeqar
characterized by the five free parameters,
\beq
\delta_Z , x_\gamma, x_Z , y_\gamma , y_Z,
\label{3.4}
\eeq
which vanish for the standard case.
We refer to the Lagrangian (3.3) as the ``phenomenological non-standard
Lagrangian'' for interactions among vector-bosons.
It has been widely used in the simulation, e.g.\  \cite{billy,grieche},
of the analysis of future data on,
e.g.,\ $e^+e^-\to W^+W^-$.

Electromagnetic gauge invariance of the Lagrangian (\ref{3.3}) and
invariance of the $Z_\mu$ field requires $W^{\pm\mu\nu}$ to transform
according to (\ref{emcharged}) and to have the non-Abelian form
\beqar
W^{\pm\mu\nu}&=&W^{\pm\mu\nu}_{(A)}
\mp ig(W^{\pm\mu}W^\nu_3 - W^\mu_3 W^{\pm\nu} ),
\label{nonab}\eeqar
where $W^\mu_3$ is to be replaced by the linear combination of the
$Z_\mu$ and $A_\mu$ following from (\ref{physfields}).
The non-Abelian structure (\ref{nonab}) induces quadrilinear interactions
among two charged and two neutral vector bosons and fixes the strength of
these interactions. Electromagnetic gauge invariance allows to add
quadrilinear interactions among four charged bosons of arbitrary strength
to (\ref{3.3}). We fix the strength of these interactions by adopting the
non-Abelian forms
\beqar
F^{\mu\nu}&=&F^{\mu\nu}_{(A)}+ie(W^{+\mu}W^{-\nu}-W^{-\mu}W^{+\nu}),\nn\\
Z^{\mu\nu}&=&Z^{\mu\nu}_{(A)}+ie\frac{c_W}{s_W}
(W^{+\mu}W^{-\nu}-W^{-\mu}W^{+\nu})
\eeqar
for $F^{\mu\nu}$ and $Z^{\mu\nu}$ in (\ref{3.3}) which with
(\ref{physfields})
supplement the tensors (\ref{nonab}) by their neutral non-Abelian
component $W^{\mu\nu}_3$ and by $B^{0\mu\nu}$.

In order to explicitly display and discuss the
$SU(2)_L \times U(1)_Y$-symmetry properties of the non-standard
interactions
in (\ref{3.3}), we transform to the $W^3B^0$ base \cite{billy}
and the matrix notation already employed in (\ref{kin}),
(\ref{2.5}). Applying the transformation
(\ref{physfields}), we obtain
\beqar
{\cal L}_{\mathrm{int, NS}}&=&2\,c_1\,{\mathrm{tr}}\,[(W^\mu
-\frac{g'}{g}B^\mu)
W_{\mu\nu}(W^\nu-\frac{g'}{g}B^\nu)]\nn\\
&&+2\,c_2\,{\mathrm{tr}}\,[B_{\mu\nu}W^\mu W^\nu ]\nn\\
&&+c_3\,{\mathrm{tr}}\,
[\tau_3\,W_{\mu\nu}]\,{\mathrm{tr}}\,[\tau_3\,W^\mu W^\nu]\nn\\
&&+\frac{c_4}{M_W^2}\,{\mathrm{tr}}\,
[W_\mu^{~\nu}W_\nu^{~\lambda}W_\lambda^{~\mu}]\nn\\
&&+2\,\frac{c_5}{M_W^2}\,{\mathrm{tr}}\,[B_\mu^{~\nu}W_\nu^{~\lambda}
W_\lambda^{~\mu}],
\label{3.8}
\eeqar
where the five coefficients are linearly related to the five parameters
in (\ref{3.3}) via
\beqar
c_1 & = & ie c_W \delta_Z , \nn \\
c_2 & = & -ie ( c_W x_\gamma - s_W x_Z- s_W \delta_Z ), \nn \\
c_3 & = & -ie (c_W x_Z + s_W x_\gamma ) , \nn \\
c_4 & = & -\frac{2}{3}\,i\,e\,(c_W y_Z + s_W y_\gamma ), \nn \\
c_5 & = & ie (s_W\,y_Z - c_W\,y_\gamma ).
\label{3.7}
\eeqar
The full vector-boson Lagrangian is obtained by adding (\ref{3.8})
to the sum of (\ref{kin}) and (\ref{2.5}).
Having established electromagnetic gauge invariance of the Lagrangian
(\ref{3.3}), and consequently of (\ref{3.8}),
the various steps of the previous section may now be applied
to this extended Lagrangian.

Requiring $SU(2)_L \times U(1)_Y$ symmetry to be realized by
the vector-boson interactions themselves (step i)) immediately excludes
all non-standard terms in (\ref{3.8}) with the only exception of
the term with coefficient $c_4 \not= 0$, i.e.,
\beq
c_1 = c_2 = c_3 = c_5 = 0 .
\label{3.9}
\eeq
We remain with a single non-vanishing independent parameter.
{}From the explicit expressions of the coefficients in (\ref{3.7})
one immediately finds $\delta_Z=x_\gamma=x_Z=0$ and the relation
\beq
y_Z = \frac{c_W}{s_W} y_\gamma
\label{3.10}
\eeq
first given in \cite{krs}. Note
that (\ref{3.10}) generalizes the relation $g_{ZWW} = e\,(c_W / s_W)$
in (\ref{3.2}) to the quadrupole interaction.

We turn to exclusion of intrinsic $SU(2)$ violation \cite{kmss},
i.e., $SU(2)_C$ symmetry (step ii)). This requirement is much less restrictive,
as it demands a vanishing value of only\footnote{Note that the decoupling
limit
$g^\prime = e = s_W=0$ of custodial $SU(2)$ symmetry in (\ref{3.8})
has to be taken under the constraints $e/s_W = {\mathrm{const}}$ as
well as $c_3 = {\mathrm{const}}$ and $c_4 = {\mathrm{const}}$
(compare \cite{billy}).}
\beq
c_3 = 0,
\label{3.11}
\eeq
in (\ref{3.8}), while the interactions involving $B_\mu$ and $B_{\mu\nu}$ are
allowed in analogy to the symmetry property of the standard
vector-boson mass term (\ref{2.5}). The requirement (\ref{3.11})
implies
\beq
x_Z =  -\frac{s_W}{c_W} x_\gamma ,
\label{3.12}
\eeq
a relation first given in \cite{kmss}. Accordingly, upon imposing $SU(2)_C$
symmetry we remain with two independent parameters $(\delta_Z , x_\gamma)$ in
the dimension-four part of the Lagrangian (\ref{3.3}), (\ref{3.8}),
and with the additional two independent parameters $(y_\gamma , y_Z)$ in the
dimension-six part. Imposing the symmetry restriction
(\ref{3.10}) on the dimension-six terms, i.e., combining (\ref{3.10})
with (\ref{3.12}), leads to a Lagrangian with three independent parameters
$(\delta_Z , x_\gamma , y_\gamma)$
which embodies $SU(2)_C$ in the dimension-four
and local $SU(2)_L \times U(1)_Y$ in the dimension-six terms.

Finally, requiring $SU(2)_C$ and minimal coupling of the hypercharge field
(no $B_{\mu\nu}$ term in (\ref{3.8})), implies
$c_2=c_3=c_5=0$, i.e.
(\ref{3.10}), (\ref{3.12})
as well as \cite{bksetal}
\beq
\delta_Z=\frac{x_\gamma}{c_W\,s_W}.
\eeq
We remain with only two independent parameters, $(x_\gamma,y_\gamma)$.

Even though $SU(2)_C$ is directly tested
by the experimentally measured parameter $\Delta x^{\mathrm{exp}}$
in (\ref{2.6}) which determines the mass ratio $M_{W^0}/M_{W^\pm}$,
its extension to self couplings amounts to an assumption, after all. It is of
interest, accordingly, that this underlying assumption cannot only be tested
in multi-parameter cases but also in a single-free-parameter model. Simply
imposing $c_1 = c_2 = c_4 = c_5 = 0$ in (\ref{3.8}), (\ref{3.7})
implies the constraint
\beqar
x_Z = \frac{c_W}{s_W} x_\gamma
\label{couprel}
\eeqar
with $\delta_Z = y_\gamma = y_Z = 0$ which obviously yields $c_3 \not= 0$
and a single-free-parameter test of $SU(2)_C$ symmetry.

Local $SU(2)_L \times U(1)_Y$ symmetry of the $U(1)_{em}$
phenomenological non-standard
Lagrangian (\ref{3.3}), (\ref{3.8}) becomes manifest by applying
the Stueckelberg transformation (\ref{2.8}), (\ref{2.14}) on it (step iii)).
Noting that the $c_2$ and $c_3$ terms in (\ref{3.8})
are invariant under the substitution
$W^\mu W^\nu\to (W^\mu-\frac{g'}{g}B^\mu)(W^\nu-\frac{g'}{g}B^\nu)$
and making use of (\ref{transf}), one finds
\newpage
\beqar
{\cal L}_{\mathrm{int, NS}} & = &  \quad 2\,\frac{c_1}{g^2}\,{\mathrm{tr}}\,
[ ( D_\mu U )^\dagger
W^{\mu\nu} (D_\nu U)] \nn \\
&&+  2\,\frac{c_2}{g^2} \, {\mathrm{tr}}\,
[ B^{\mu\nu} ( D_\mu U)^\dagger (D_\nu U) ]\nn\\
&&+  \frac{c_3}{g^2} \,{\mathrm{tr}}\, [ U\,\tau_3 \,U^\dagger
\,W^{\mu\nu} ]\,
{\mathrm{tr}}\, [ \tau_3 \,(D_\mu U)^\dagger \,(D_\nu U)] \nn \\
&&+ \frac{c_4}{M^2_W} \, {\mathrm{tr}}\, [ W_\mu^{~\nu}
W_\nu^{~\lambda} W_\lambda^
{~\mu} ] \nn \\
&&+  2\,\frac{c_5}{M^2_W} \,{\mathrm{tr}}\,
[ B_\mu^{~\nu} U^\dagger W_\nu^{~\lambda}
W_\lambda^{~\mu} U].
\label{3.13}
\eeqar
We stress that the Lagrangians (\ref{3.3}), (\ref{3.8}) and (\ref{3.13})
are equivalent.
Consequently, no new arguments become available concerning the
reduction of the number of five free parameters by rewriting
(\ref{3.3}), (\ref{3.8}) in the manifestly $SU(2)_L\times U(1)_Y$
gauge invariant form (\ref{3.13}). Often, this point is not correctly
presented
in the literature. While all three dimension-four terms in (\ref{3.13}) are
given in \cite{appelwu,hewe}, the $SU(2)_C$-violating $c_3$-term is not
 present
in e.g. \cite{holdom}, and the omission is not justified.
The form (\ref{3.13}) of the Lagrangian is of interest as it
connects the subject of non-standard vector-boson interactions with the
widely used investigations (e.g.~\cite{pich}) on chiral Lagrangians.
Also, (\ref{3.13}) provides an intermediate step in introducing
the Higgs scalar.

In the last step (step iv)), we now assume the existence of a Higgs
scalar particle of mass $m_H$ sufficiently below the unitarity limit of
$m_H \simeq 1$ TeV.
Carrying out the substitution (\ref{2.13}) in (\ref{3.13}), one obtains
\beqar
{\cal L}_{\mathrm{int, NS}} & = & \quad\frac{c_1}{M_W^2} \, {\mathrm{tr}}\,
[ ( D_\mu \phi)^
\dagger W^{\mu\nu} (D_\nu \phi )] \nn \\
&&+  \frac{c_2}{M^2_W} \, {\mathrm{tr}}\, [ B^{\mu\nu} (D_\mu \phi)^
\dagger (D_\nu \phi)] \nn \\
&&+  \frac{c_3}{M_W^4}\,
\frac{g^2}{4}\,{\mathrm{tr}}\, [ \phi\,\tau_3 \,\phi^\dagger\,
W^{\mu\nu} \phi] \, {\mathrm{tr}}\,
[ \tau_3 \,(D_\mu \phi)^\dagger \,(D_\nu \phi)]
\nn \\
&&+  \frac{c_4}{M^2_W} \, {\mathrm{tr}}\, [ W_\mu^{~\nu}
W_\nu^{~\lambda} W_\lambda^
{~\mu} ]  \nn \\
&&+   \frac{c_5}{M_W^4}\, g^2 \,{\mathrm{tr}}\, [ B_\mu^{~\nu} \phi^
\dagger W_\nu^{~\lambda} W_\lambda^{~\mu} \phi ].
\label{3.14}
\eeqar
As far as vector-boson interactions are concerned, (\ref{3.14})
is equivalent to the phenomenological non-standard Lagrangian
(\ref{3.3}). The linearization of the scalar sector and the
introduction of a
physical Higgs boson does not change the
vector boson self-interactions.

The introduction of the Higgs particle, when passing from (\ref{3.13})
to (\ref{3.14}), leads to a dimensional transmutation of the
interaction terms. The Lagrangian (\ref{3.14}) contains three dimension-six
and two dimension-eight terms.
Restricting oneself to dimension-six terms \cite{deru,hagiwara,hewe}
in (\ref{3.14}) implies
\beq
c_3=c_5=0 ,
\eeq
and thus the relations (\ref{3.10}) and (\ref{3.12}) \cite{dreimann}
originally derived from symmetry arguments \cite{kmss,krs}.

The omission of the higher-dimension terms in (\ref{3.14})
may be based on the expectation that such terms
would lead to a less decent high-energy behavior
of scattering cross sections than terms of lower dimension.
This is not
the case, however, for the $c_3$ and $c_5$ terms in (\ref{3.14}).
In fact, all terms in the Lagrangian (\ref{3.14}) give rise to the same
high-energy behavior of boson-boson scattering cross sections. To be
specific, {\em as long as terms quadratic in the deviations
from the standard model are neglected}, the cross sections for all
boson-boson
scattering processes become constant\footnote{Remember that in the
standard model all cross sections are ${\cal O}(s^{-1})$.} (i.e. they
are ${\cal O}(s^0)$, where
$\sqrt{s}$ denotes the c.m. scattering energy) in the high-energy limit.
This is true for all terms in (\ref{3.14})\footnote{If
the Higgs boson is omitted from the theory, i.e. if one uses the sum
of the Lagrangians (\ref{kin}), (\ref{2.10}) and (\ref{3.13}), the cross
sections rise as ${\cal O}(s)$ independently of whether only $c_1$, $c_2$
and $c_4$ or also $c_3$ and $c_5$ are different from zero (if one
neglects the
terms quadratic in the non-standard couplings).
The less decent high energy behavior as compared to the case with a
Higgs boson
is entirely due to the omission of the {\em standard\/} Higgs interactions
\cite{et,kuss}.}.
{}For the scattering of two vector-bosons
into two vector-bosons the ${\cal O}(s^0)$ behavior for all terms
has been shown in \cite{et}. Using the formalism of \cite{et}, one easily
generalizes this result of a constant, ${\cal O}(s^0)$, behavior to all
two-boson into two-boson processes, i.e.~for any combination of
external Higgs bosons and vector-bosons in an arbitrary polarization state.
Accordingly, the omission of the $c_3$ and $c_5$ terms
is of no relevance for the high-energy behavior of scattering processes.
The dimensional argument cannot be justified by a reference to
differences in the high-energy behavior.

As no direct empirical evidence for the
Higgs particle is available so far, it is gratifying that the
three-parameter model of Lagrangian (\ref{3.3}), (\ref{3.8})
with the constraints
(\ref{3.10}) and (\ref{3.12}) does not rely on the existence of a
Higgs particle. Simple symmetry considerations alone are sufficient to
exclude theoretically less favored scenarios.

Let us note,
however, that the (linearly realized) $SU(2)_L\times U(1)_Y$ symmetry
according to (\ref{3.14}) specifies certain non-standard Higgs
interactions, in case a light Higgs particle is realized in nature in
conjunction with non-standard vector-boson self-couplings.
The introduction of the Higgs particle improves the high-energy behavior
of the standard electroweak theory from an ${\cal O}(s^0)$ to an
${\cal O}(s^{-1})$ behavior. As a consequence, also the high-energy
behavior
of the theory with non-standard couplings is improved
from ${\cal O}(s)$ to ${\cal O}(s^0)$, as long as the non-standard
parameters
are small enough to be treated in linear approximation.
The introduction of the non-standard Higgs interactions does not provide
an
additional
argument, however, for omitting certain non-standard terms in
(\ref{3.3}), (\ref{3.8}), (\ref{3.13}),
apart from those arguments already derived from symmetry
considerations.

\section{Conclusion}
Various authors have independently derived the two aforementioned
$SU(2)_L\times U(1)_Y$-gauge-invariant Lagrangians in a chiral
Lagrangian approach and within an effective field theory with a light
Higgs particle. Dimensional arguments were used to reduce the number
of independent parameters. In the present note we have shown that the
different Lagrangians emerge from a phenomenological Ansatz
by applying a Stueckelberg transformation
followed by a linearization of the scalar sector.
The results of dimensional arguments concerning the reduction of the number
of free parameters when applied to the Lagrangian with the Higgs scalar
coincide with the result of the symmetry arguments
suggested and employed a long time ago
in the framework of the phenomenological Lagrangian.
The dimensional arguments, however, cannot be justified
from the high-energy
behavior, as all five non-standard terms lead to the same
behavior.

\section*{Acknowledgement}
The authors thank S. Dittmaier, G. Gounaris and G. Weiglein for
useful discussions.

\end{document}